\newfont{\msbm}{msbm10}

\newfont{\msbms}{msbm6}  

\def\a{\alpha}
\def\b{\beta}
\def\ba{\begin{eqnarray}}
\def\ea{\end{eqnarray}}
\def\be{\begin{equation}}
\def\ee{\end{equation}}

\documentclass[aps,prd,showpacs,preprint,amssymb,nofootinbib]{revtex4}

\usepackage{amssymb,amsmath}
\usepackage{amsfonts}
\usepackage{mathrsfs}
\usepackage{txfonts}
\usepackage{marvosym}
\usepackage{amssymb}
\usepackage{color}

\begin{document}


\title[Quantum  unitary dynamics in cosmological spacetimes]{Quantum  unitary dynamics in cosmological spacetimes}

\author{Jer\'onimo Cortez}
\affiliation{Departamento de F\'\i sica,
Facultad de Ciencias, Universidad Nacional Aut\'onoma de
M\'exico, M\'exico D.F. 04510, Mexico.}
\email{jacq@ciencias.unam.mx}
\author{Guillermo A. Mena
Marug\'an}
\affiliation{Instituto de Estructura de la Materia, IEM-CSIC,
Serrano 121, 28006 Madrid, Spain.}
\email{mena@iem.cfmac.csic.es}
\author{Jos\'e M. Velhinho}
\affiliation{Departamento de F\'{\i}sica, Faculdade de Ci\^encias, Universidade
da Beira Interior, R. Marqu\^es D'\'Avila e Bolama,
6201-001 Covilh\~a, Portugal.}
\email{jvelhi@ubi.pt}

\begin{abstract}
We address the question of unitary implementation of the dynamics for scalar fields in cosmological scenarios. Together with invariance under spatial isometries, the requirement of a unitary evolution singles out a rescaling of the scalar field and a unitary equivalence class of Fock representations for the associated canonical commutation relations. Moreover, this criterion provides as well a privileged quantization for the unscaled field, even though the associated dynamics is not unitarily implementable in that case. We discuss the relation between the initial data that determine the Fock representations in the rescaled and unscaled descriptions, and clarify that the S-matrix is well defined in both cases. In our discussion, we also comment on a recently proposed generalized notion of unitary implementation of the dynamics, making clear the difference with the standard unitarity criterion and showing that the two approaches are not equivalent.
\end{abstract}

\pacs{04.62.+v, 98.80.Qc}

\maketitle

\section{Introduction}
\label{intro}

In a series of works completed during the last ten years \cite{9,10,15,16,CQG28,21}, strong results have been obtained by us and other collaborators concerning the Fock quantization of scalar fields with an effective time dependent mass and the preservation of unitarity. In particular, when applied to the quantization of free (test) fields in an expanding background spacetime with homogeneous and isotropic spatial sections [i.e., a Friedmann-Lema\^{\i}tre-Robertson-Walker (FLRW) spacetime], these results allow one to select a privileged parametrization of the field variable, that involves a rescaling of the original field by means of the time dependent scale factor of the background \cite{18,21,22}. 

For definiteness, let us consider the case of a flat FLRW spacetime, with compact spatial sections isomorphic to the 3-torus. Using conformal time, the metric can be written in the form
\be
ds^2=a^2(\eta)\left(-d\eta^2+d{\vec x}^{\,2}\right),
\ee
where $a(\eta)$ is the scale factor and $d{\vec x}^{\,2}$ is the standard metric on the 3-torus. It is well known that the free scalar field equation
\begin{equation}
\label{wave}
\left(\Box -m^2\right)\phi=0,
\end{equation}
acquires its simplest form after the change of field variable
\begin{equation}
\label{rescale}
\chi=a \phi,
\end{equation}
leading to 
\begin{equation}
\label{feq}
{\chi}''
-\Delta\chi +m^2(\eta)\chi=0,
\end{equation}
where $m^2(\eta)=a^2m^2-(a''/a)$ and $\Delta$ is the standard Laplacian on the 3-torus. Besides, the box stands for the D'Alambertian of the FLRW spacetime, and the derivative with respect to the conformal time is denoted with a prime.

For the type of equation (\ref{feq}), it is known that there is a quantum Fock representation of the canonical commutation relations (CCR's) at fixed (equal) time such that the dynamics in the canonical framework is unitarily implementable. Moreover, additional research \cite{16,18,21,22}, in which we participated, has proven that this quantization is unique [among the set of Fock quantizations determined by complex structures (CS's) that are invariant under the action of spatial isometries]. It has also been shown that unitary implementation of the dynamics is impossible to achieve for the type of equations satisfied by the original unscaled field $\phi$, or by means of any other rescaling different from the one introduced above: $\phi \to \chi=a\phi$.

The usefulness of the above transformation, $\phi \to \chi$, in particular to simplify the field equation, has of course been known for a long time (see e.g. \cite{Fb}) and it is commonly used, even in the anisotropic cosmology context (see for instance \cite{BD}). Moreover, nothing is lost by considering the field $\chi$, since the construction of field operators $\hat \chi(\eta)$ in the quantization process immediately gives rise to operators $\hat \phi(\eta)=a^{-1}(\eta)\hat \chi(\eta)$ to represent the original field. However, there is an apparent tension between the two formulations, in the light of the different properties of the dynamics in each of them. In order to try to alliviate this tension, a recent proposal of a generalized notion of unitary implementation of the dynamics has been put forward in \cite{AA}, to accomodate the fact that evolution in terms of the $\phi$-field description is not unitary.

Nonetheless, in our opinion this tension is somewhat artificial. The main purpose of this article is to clarify this issue, explaining how one can mantain a well-defined and non-trivial concept of evolution in the quantization, and how this concept can be employed 
as a selection criterion for the determination of a unique Fock quantization. The rest of the paper is organized as follows. We summarize the basic mathematical tools needed to discuss the role of the dynamics in the Fock quantization of a scalar field in Sec. II. Sec. III deals with the unitary implementability of the dynamics and the determination of criteria that select a unique Fock representation for the quantization of the scalar field. Sec. IV analyzes the relation between the quantization of the rescaled field and the original one, and how this rescaling affects the unitarity of the dynamics. Finally, we present a discussion of the implications of our investigations and conclude in Sec. V.

\section{Quantum field theory in cosmology}

In the traditional Hilbert space based approach (in contrast with the algebraic or representation independent approach), a major question in quantum field theory (QFT) is the selection of an appropriate representation of the CCR's, or the construction of appropriate quantum field operators in a given Hilbert space satisfying the wave equation. In fact, the fundamental Stone-von Neumman uniqueness theorem has no correspondent in field theory. Let us recall that, for finitely many degrees of freedom, the Stone-von Neumann theorem states that all quantizations satisfying appropriate continuity conditions are unitarily equivalent. However, in field theory, one has to deal with infinitely many, not unitarily equivalent quantizations \cite{wald}. For linear theories, one has at least a good understanding of the problem, both from the covariant and the canonical perspectives, and the freedom in the construction of a representation can be put in terms of CS's. A CS is a real linear transformation that leaves invariant the symplectic structure of the system (and hence its CCR's) and with a square equal to minus the identity (the attention is focused just on compatible CS's, that provide an inner product on the one-particle Hilbert space when they are suitably combined with the symplectic structure \cite{wald}).
 
The usual Fock quantizations of a given linear scalar field are completely determined by CS's in the space of solutions $\Gamma_{\rm Cov}$ of the corresponding wave equation, in our case equation (\ref{wave}), or (\ref{feq}), depending on the chosen variable. For a thorough discussion of Fock representations, in particular adapted to scalar fields in FLRW spacetimes, we refer the reader to \cite{adia} (we follow also  \cite{AA} in what concerns the relation between the covariant and the canonical descriptions). Of course, since solutions are determined by Cauchy data at any given time, CS's in the space of solutions give rise to CS's in the canonical phase space. To adopt a similar notation to that employed in \cite{AA}, let us fix a 3-manifold $M$, homeomorphic to the 3-torus. The canonical phase space $\Gamma_{\rm Can}$ can be viewed as the  set of pairs of functions on $M$ with the usual symplectic structure (each pair of functions supplying the values of the field and its momentum). Now, for each time $\eta$ there is a (symplectic) isomorphism $I_{\eta}:\Gamma_{\rm Cov}\to \Gamma_{\rm Can}$, given by
\be
I_{\eta}\phi(\vec x,\eta)=\left(\varphi(\vec x),\pi_{\varphi}(\vec x)\right),
\ee
where
\be
\varphi(\vec x)=\phi(\vec x,\eta);\ \ \pi_{\varphi}(\vec x)=a^2(\eta) \frac{\partial\phi}{\partial\eta}(\vec x,\eta).
\ee
Of course, the same is true concerning the field equation (\ref{feq}), with the difference that the canonical momentum is now given by $\pi_{\chi}=\partial\chi/\partial\eta$.

Suppose  that we are given a  CS $\cal J$ in $\Gamma_{\rm Cov}$. Then, the maps $I_{\eta}$ induce a 1-parameter family of CS's in the canonical phase space  $\Gamma_{\rm Can}$, by
\be
\label{7}
J_{\eta}=I_{\eta} {\cal J} I^{-1}_{\eta}.
\ee
For any given $\cal J$, this family of CS's in $\Gamma_{\rm Can}$ satisfy, by construction, the following relations:
\be
\label{77}
J_{\eta_2}=E_{\eta_2,\eta_1} J_{\eta_1} E^{-1}_{\eta_2,\eta_1},\ee
for all pairs of instants of time (i.e., $\forall \eta_1,\eta_2$,), and where
\be
\label{evol}
E_{\eta_2,\eta_1}=I_{\eta_2}  I^{-1}_{\eta_1}: \Gamma_{\rm Can}\to\Gamma_{\rm Can}.
\ee
Conversely, a CS $\cal J$ in $\Gamma_{\rm Cov}$ is determined by a family of CS's in $\Gamma_{\rm Can}$ satisfying (\ref{77}).

Of course, one can fix once and for all a given time $\eta_0$ in order to identify $\Gamma_{\rm Cov}$ with $\Gamma_{\rm Can}$. From this perspective, the CS
\be
\label{jev}
J_{\eta}=E_{\eta,\eta_0} J_{0} E^{-1}_{\eta,\eta_0},
\ee  
where
\be
J_{0}=I_{\eta_0} {\cal J} I^{-1}_{\eta_0},
\ee  
is the CS generated by dynamical evolution of $J_0$ (at initial time $\eta_0$), since the maps $E_{\eta_2,\eta_1}$ clearly describe the time evolution in canonical phase space.

A CS $J_0$ in $\Gamma_{\rm Can}$ determines a representation of the CCR's (at equal time). It is clear that a unitary implementation of the dynamics exists in the representation determined by $J_0$ if and only if the representations of the CCR's determined by the transformed CS's $J_{\eta}$ (\ref{jev}) are unitarily equivalent to the one determined by $J_0$. This in turn is true if and only if the operator $J_{\eta}-J_0$ is Hilbert-Schmidt (on the one-particle Hilbert space of the Fock representation given by $J_0$). Recall that an operator is Hilbert-Schdmit if the product of its adjoint with itself is trace-class.

Of course, the notion of unitary time evolution may be formulated within the covariant framework as well. In fact, once a reference time $\eta_0$ has been fixed, ${\cal J}= I^{-1}_{\eta_0}JI_{\eta_0}$ establishes a one-to-one correspondence  between CS's in $\Gamma_{\rm Can}$ and in $\Gamma_{\rm Cov}$,and therefore the family of CS's $J_{\eta}$ (\ref{jev}) provides a family ${\cal J_{\eta}}= I^{-1}_{\eta_0}J_{\eta}I_{\eta_0}$ of CS's in $\Gamma_{\rm Cov}$. The above unitarity condition on the dynamics translates into the requirement that ${\cal J}_{\eta}-{\cal J}$ be Hilbert-Schmidt ($\forall \eta$), a requirement which amounts to the unitary implementation of the transformations that describe the time evolution in $\Gamma_{\rm Cov}$, namely, ${\cal E}_{\eta \eta_0}=I^{-1}_{\eta_0}E_{\eta \eta_0}I_{\eta_0}=I^{-1}_{\eta_0}I_{\eta}$.

To conclude this overview let us make contact with the standard textbook treatments of QFT in curved spacetime (CST), where one typically deals with the covariant formulation in terms of so-called mode solutions. Here, one writes the quantum field in terms of (e.g.) Fourier modes
\be
\label{qf}
\hat\phi(\vec x,\eta)=\frac{1}{(2\pi)^{3/2}a(\eta)}\sum_{\vec k}\left[\chi_{k}\hat A_{\vec k}+
\chi^*_{k}\hat A^{\dagger}_{-\vec k}\right]e^{i\vec k\cdot\vec x},
\ee
where $k=|\vec k|$, 
the symbol $*$ denotes the complex conjugate, and the dagger stands for the adjoint of the corresponding operator. The above expression for the field defines in fact operator valued distributions; obtaining {\it bona fide} operators requires a smearing with appropriate test functions on the spatial manifold 
$M$:
\be
\label{qg}
\hat\phi(f,\eta)=\int_Mf(\vec x)\hat\phi(\vec x,\eta)d^3x= \frac{1}{a(\eta)}
\sum_{\vec k}{f_{-\vec k}\left[\chi_{k}\hat A_{\vec k}+
\chi^*_{k}\hat A^{\dagger}_{-\vec k}\right],}
\ee 
where $f_{\vec k}=(2\pi)^{3/2}\int f \exp(-i \vec{k}\cdot \vec{x})$ denotes the Fourier coefficient of  the test function $f(\vec x)$ (one can choose
functions $f$ such that the set of non-zero coefficients $f_{\vec k}$ is finite, for simplicity). 

We have chosen to explicitly factor out the inverse of the scale factor in (\ref{qf}), which is a common practice but by no means necessary. Here, the functions $\chi_k(\eta)$ are solutions to the equations of motion that follow from the field equation (\ref{feq}) for each of the Fourier modes (of course, different modes $\phi_k=a^{-1}\chi_k$ obey different equations of motion):
\be
\chi_k''+\left(k^2+m^2(\eta)\right)\chi_k=0.
\ee
The normalization condition 
\be
\label{norm}
\chi_k\chi'^*_k-\chi_k'\chi^*_k=i
\ee
(which is time-independent) ensures that the operators $\hat A_{\vec k}$ and $\hat A^{\dagger}_{\vec k}$ satisfy commutation relations of creation-annihilation type.

The Fock quantization is determined by declaring that   $\hat A_{\vec k}$ and $\hat A^{\dagger}_{\vec k}$ are the annihilation-creation operators for each mode. Different Fock quantizations are obtained by changing the quantization maps that assign a meaning to the operators  $\hat A_{\vec k}$ and $\hat A^{\dagger}_{\vec k}$, something which is effectively done by changing the mode solutions $\chi_k$. A given set of mode solutions is obtained, for instance, by fixing initial data for $\chi_k$ and $\chi'_k$ at a given time, which in turn determines exactly which (complex) classical variables are being represented by the operators $\hat A_{\vec k}$ and $\hat A^{\dagger}_{\vec k}$. The reader can consult \cite{adia} for the explicit relation between initial data for the modes and CS's,
both in the covariant and in the canonical approach.

\section{Selection criteria}

How can one then select a CS adequate to a given linear scalar field theory? Notice first that, in the general case, there is no reason to expect that a unique CS will stand out, neither that is necessary: for the unambiguous construction of a Hilbert space based quantum theory, it is sufficient that one can identify a preferred unitary equivalence class of representations.

The simplest situation is, of course, that of the free field in Minkowski spacetime. It turns out that, for a given value of the mass, there is a unique CS $\cal J$ which is invariant under the natural action of the Poincar\'e group on the space of solutions. This unique CS corresponds, indeed, to the usual creation-annihilation operators. In terms of initial data for the Fourier modes, it is determined by the choice
\be
\phi_k=\frac{1}{\sqrt{2\omega}}, \qquad \phi'_k=-i\sqrt{\frac{\omega}{2}}
\ee
(up to a phase) at a given instant, with
\be
\omega^2 =k^2+m^2.
\ee
The invariance of the CS under the Poincar\'e group naturally gives rise to a unitary implementation of the group, which includes time translations. From the canonical perspective, this means that the CS is invariant under time evolution. Classical dynamics are therefore unitarily implemented, along with the remaining set of isometries, in the most natural way. From a broad perspective, the lesson is that, in field theory, to fix the quantum representation, one must use not only the kinematical setup, but also the dynamics and the set of symmetries.  From the point of view of QFT in CST, one may also say that the  properties of the spacetime itself enter the specification of the quantum theory. 

In general terms, Fock quantizations can be defined for any linear symplectic space, and a unique quantization is singled out by the following requirement \cite{kay, BSZ, AM1,AM2}. {Suppose that one has a distinguished one-parameter group of symplectic transformations (in favourable cases, of course, this group is generated by the Hamiltonian). Then, there exists a unique CS which is invariant under the action of this group and such that the generator of the corresponding one-parameter unitary group (the existence of which is guaranteed by the invariance of the CS) is a positive operator on the Fock space.} The issue in QFT in CST is that, in general, there is no distinguished family of one-parameter symplectic transformations that one can use, except in the stationary situation, i.e., when a timelike Killing vector field is present. 
 
One possible way around this lack of uniqueness in the quantization of fields in CST is to try and replace the requirement of invariance under dynamical evolution by a weaker one, requiring only a unitary implementation of the dynamics. In technical terms, this boils down to abandon the idea of an invariant CS, which is not viable, in favour of a CS which transforms, under time evolution, into a CS that belongs to the same unitary equivalence class. This still gives a unitary implementation of the dynamics in a Fock representation, with the difference that one no longer requires the existence of a {state} which is invariant under evolution. This is the approach adopted in our works \cite{9,10,15,16,18,21,22}. Different possibilities were explored since the 1970's in order to remove the ambiguity in QFT in CST (see  e.g. \cite{Fb,adia,jackiw}), and we would like now to compare the most employed of those approaches with ours\footnote{A different route, which in part emerged from the limitations that the mentioned approaches necessarily present, is to abandon the idea of fixing quantum representations altogether, and adopt a representation independent version of QFT, along the lines of algebraic QFT. We will not explore this avenue here.}.

To get some insight into the problem and better motivate our proposal, let us consider first the case of a spacetime which, for $\eta$ in the past of some fixed time $\eta_1$, is isomorphic to {(a part of)} Minkowski spacetime. In this situation, it seems clear that an unambiguous and physically sensible way to proceed is to perform a canonical quantization at a fixed time $\eta<\eta_1$, using the same CS in canonical phase space as if we were in presence of a free field in actual Minkowski spacetime. Notice that for $\eta<\eta_1$, when we have that $a(\eta)$ is a constant, $a_{in}$, the equations of motion for the modes (of the field $\phi$) are 
\be
\label{20}
\phi_k''+\left(k^2+m^2a^2_{in}\right)\phi_k=0.
\ee
In terms of initial data for the Fourier modes, the proposed quantization corresponds to the choice
\be
\label{modein}
\phi_k=\frac{1}{a_{in}\sqrt{2\omega}}, \qquad \phi'_k=-\frac{i}{a_{in}}\sqrt{\frac{\omega}{2}}
\ee
at a fixed instant $\eta<\eta_1$, with
\be
\label{ome}
\omega^2 =k^2+m^2a^2_{in}.
\ee
The initial data are thus, in fact, those corresponding to a free field, with the {\it local} value of the mass $m^2(\eta)=m^2a^2_{in}$, and normalized such that (\ref{norm}) is satisfied.

A situation of great conceptual interest occurs when the spacetime is isomorphic to Minkowski spacetime both in the distant past and in the far future. In our current {homogeneous} and isotropic cosmology setup this happens if, besides $\eta_1$ as above, there is also an instant of time $\eta_2$ such that $a(\eta)$ is constant $\forall \eta>\eta_2$ (the constant values of the scale factor in the far past and future, $a_{in}$ and $a_{out}$, need not be the same).

One can now apply the procedure explained above both at $\eta<\eta_1$ and at $\eta^\prime>\eta_2$, obtaining two distinct CS's in $\Gamma_{\rm Cov}$, say ${\cal J}_{in}$ defined by (\ref{modein}) and (\ref{ome}), and ${\cal J}_{out}$ defined by 
\be
\label{modeout}
\phi_k=\frac{1}{a_{out}\sqrt{2\omega'}}, \qquad \phi'_k=-\frac{i}{a_{out}}\sqrt{\frac{\omega'}{2}},
\ee
\be
\label{ome2}
\omega'^2 =k^2+m^2a^2_{out},
\ee
at a fixed instant $\eta'>\eta_2$. If the CS's ${\cal J}_{in}$ and ${\cal J}_{out}$ were to give rise to quantum theories which are not unitarily equivalent, i.e. if ${\cal J}_{in}-{\cal J}_{out}$ were not Hilbert-Schmidt, we would have to face a serious ambiguity in the quantization procedure. Fortunately this is not the case. It has been established for a long time that, in this situation, ${\cal J}_{in}-{\cal J}_{out}$ is a Hilbert-Schmidt operator. In fact, this is a particular case of the well established result about the existence of the S-matrix (see \cite{W,D}). Moreover, this has been proven directly, see e.g. \cite{Fb,Fp}.

At this stage of our discussion, it is worth commenting on certain subtleties related with the equivalence of the two CS's ${\cal J}_{in}$ and ${\cal J}_{out}$ in the space of solutions. First, let us point out that the fact that ${\cal J}_{in}-{\cal J}_{out}$ is Hilbert-Schmidt is precisely the content of Lemma 2 in \cite{AA}. Since the authors of that work use the notation
\be
J_{in}=I_{\eta_1}{\cal J}_{in}I^{-1}_{\eta_1},
\ee
\be
J_{out}=I_{\eta_2}{\cal J}_{out}I^{-1}_{\eta_2},
\ee
it may give the impression, e.g. from Lemma 1 in \cite{AA}, that the {\sl{in}} and {\sl{out}} representations are not equivalent. However, what happens is that the evaluation of the operator $J_{out}-J_{in}$, as done in the referred Lemma 1, tell us instead about the relation between two CS's in $\Gamma_{\rm Can}$ defined by different Cauchy data at {\em the same instant of time}.

In addition to the above, to understand better the extent of the obstruction to the unitary implementation of the dynamics, even when the S-matrix is known to be well defined, let us now consider the {\sl{in}} and {{\sl out}} representations, but in terms of the field $\chi=a\phi$. We call ${\cal J}^{\chi}_{in}$ and ${\cal J}^{\chi}_{out}$, respectively, the {{\sl in}} and {{\sl out}} CS's in the corresponding space of solutions. If one defines the quantum field $\hat \chi$ using the Cauchy data that follow from (\ref{modein}) and (\ref{modeout}), namely
\be
\chi_k=\frac{1}{\sqrt{2\omega}}, \qquad \chi'_k={-i}\sqrt{\frac{\omega}{2}}, \qquad 
\omega^2 =k^2+m^2a_{in}^2
\ee
at a fixed instant $\eta<\eta_1$, and 
\be
\chi_k=\frac{1}{\sqrt{2\omega'}}, \qquad \chi'_k={-i}\sqrt{\frac{\omega'}{2}}, \qquad 
\omega'^2 =k^2+m^2a_{out}^2
\ee 
at a fixed instant $\eta'>\eta_2$, one must reach the same conclusion as before, i.e., the S-matrix exists, or equivalently ${\cal J}^{\chi}_{in}-{\cal J}^{\chi}_{out}$ is  Hilbert-Schmidt. However, there is now an important difference: precisely because the scale factor is included in the definition of $\chi$, that obeys the normalization condition (\ref{norm}), the two CS's $I_{\eta_1}{\cal J}^{\chi}_{in}I^{-1}_{\eta_1}$ and $I_{\eta_2}{\cal J}^{\chi}_{out}I^{-1}_{\eta_2}$ (in $\Gamma_{\rm{Can}}$) turn out to be equivalent 
(for compact spatial manifolds, see e.g. \cite{9}). Thus, one concludes that $I_{\eta_1}{\cal J}^{\chi}_{in}I^{-1}_{\eta_1}- E_{\eta_2,\eta_1}\left(I_{\eta_1}{\cal J}^{\chi}_{in}I^{-1}_{\eta_1}\right)E^{-1}_{\eta_2,\eta_1}$ is Hilbert-Schmidt, what is the same as saying that the time evolution is unitarily implementable with respect to the representation of the CCR's defined by  $I_{\eta_1}{\cal J}^{\chi}_{in}I^{-1}_{\eta_1}$.

After these remarks (which clarify some issues discussed recently in \cite{AA}), let us finally consider the general case of an arbitrary scale factor $a(\eta)$. Since the strategy of fixing the quantum representation by using the value of the effective mass at a given instant of time was seen to work well in the previous cases, one might expect that, at least for a slow variation of the scale factor, the same procedure could be applied without ambiguities in the general case as well. Moreover, since, for the issue of unitary equivalence (in the present context of compact spatial manifolds), only the ultraviolet modes are relevant, one might even hope that the strategy would work for arbitrary $a(\eta)$, since any variation rate, at fixed time, is small compared with arbitrarily high frequencies. It turns out that the question is not so obvious, and that the outcome heavily depends on the exact implementation of the quantization.

Let us see this in some detail. The seemingly obvious generalization of the above procedure is to define the quantization by the initial data  (\ref{modein}) and (\ref{ome}), at a given fixed time $\eta$. Let us denote the corresponding CS in $\Gamma_{\rm{Cov}}$ by ${\cal J}_{\eta}$. Since there is no preferred instant of time, one can of course apply the same procedure at any other time $\eta'$, obtaining a new CS    ${\cal J}_{\eta'}$, and therefore a new Fock quantization. It turns out that these Fock quantizations are not unitarily equivalent, i.e. ${\cal J}_{\eta'}-{\cal J}_{\eta}$ is not Hilbert-Schmidt (see e.g. \cite{Fp}), no matter how slow the variation of $a(\eta)$ is. These results were originally obtained by Parker \cite{Par1,Par2}, and were historically seen as a clear indication of serious problems with the particle interpretation of the field theory. In fact, based on the notion of particles introduced by the choice of initial data for the modes corresponding to the instantaneous value of the mass, there is a direct relation between the operator ${\cal J}_{\eta'}-{\cal J}_{\eta}$ and the number of particles in the Fock vacuum associated with ${\cal J}_{\eta'}$ with  respect to the Fock vacuum associated with ${\cal J}_{\eta}$. Again in terms of the modes, if $\hat A_{\vec k}$ and $\hat A^{\dagger}_{\vec k}$ ($\hat A'_{\vec k}$ and $\hat A'^{\dagger}_{\vec k}$) are the annihilation-creation operators corresponding to  ${\cal J}_{\eta}$ (respectively ${\cal J}_{\eta'}$), one then has a Bogoliubov transformation between them:
\be
\label{bog}
\hat A'_{\vec k}=\a_k\hat A_{\vec k}+\b_k\hat A^{\dagger}_{-\vec k},
\ee
and ${\cal J}_{\eta'}-{\cal J}_{\eta}$ is  Hilbert-Schmidt if and only if (see e.g. \cite{Fb})
\be
\sum_{\vec k}^{\infty}|\b_k|^2<\infty.
\ee

The emphasis was then put not on a particle interpretation but on an unambiguous construction of the QFT (see \cite{Fb,adia}), e.g. on the definition of a set of physically motivated  CS's (or states) which moreover give rise to unitarily equivalent quantizations\footnote{We are assuming that the spacetime has compact spatial sections, like in the particular case of the 3-torus. In the general non-compact case, the requirement of equivalence necessarily assumes a local form, since the infrared limit introduces obstructions to the unitary equivalence which are typically unavoidable. In terms of a particle interpretation, this amounts to a requirement of finite particle production in finite volume.}. A consistent answer was the set of so-called adiabatic states (we refer to \cite{adia} for a rigorous definition, including that in terms of both covariant and canonical CS's). These  states correspond to certain CS's ${\cal J}^{ad}_{\eta,r}$ in $\Gamma_{\rm Cov}$, which depend both on the instant $\eta$ at which they are defined, and on an iteration order $r$ (the superscript ${ad}$ refers to their adiabatic nature). The adiabatic states indeed fulfil the expectations, since all the associated CS's determine unitarily equivalent representations, independently of $\eta$ and of the order $r$ \cite{adia}. These results are proven in \cite{adia} for the case of FLRW with positive curvature, i.e., for the case of spatial sections isomorphic to the 3-sphere, but the arguments are immediately applicable to the 3-torus. Since we will use the result in the next section, let us see, e.g., how the independence with respect to the adiabatic order $r$ can be proven. Consider then two adiabatic states of consecutive order, associated with the CS's ${\cal J}^{ad}_{\eta,r}$ and  ${\cal J}^{ad}_{\eta,r+1}$ for some arbitrary  natural number $r\geq 0$. It is shown in \cite{adia} that the coefficient $\beta_k$ of the Bogoliubov transformation [of the type (\ref{bog})] that connects the two states behaves as $\beta_{k}=O(1/k^{2(r+1)})$ in the ultraviolet regime\footnote{The spatial sections considered in \cite{adia} are the usual ones of constant positive, negative, or zero curvature. The positive curvature case corresponds therefore to the 3-sphere, and the flat case to $\mathbb{R}^3$. The asymptotic behavior of $\beta_{k}$ is the same in all three cases, and it remains the same for the 3-torus as well. This follows immediately either from the fact that the spectrum of the Laplacian for the 3-torus is  a subset of the one for $\mathbb{R}^3$, or by realizing that the dominant ultraviolet behavior of the eigenvalues of the Laplacian is the same for both the 3-sphere and the 3-torus.}. Thus, finiteness of the sum $\sum_{\vec{k}}\vert \beta_{k}\vert^{2}$ amounts to the summability of the sequence $\{g_{k}/k^{4(r+1)}\}$, where $g_k$ is the degeneracy of each eigenspace of the Laplace operator on the 3-torus. Since, for large $k$, $g_k$ grows at most like $k^2$ (see \cite{22} for details), it follows  that $\{g_{k}/k^{4(r+1)}\}$ is indeed a summable sequence, $\forall r\geq 0$. This proves that all adiabatic  states give rise to unitarily equivalent representations, independently of their order\footnote{Actually, the sequence $\{g_{k}/w_{k}^{4}\}$ is summable for arbitrary compact Riemannian manifolds of dimension $d\leq 3$ \cite{21}, where $\{-w^{2}_k\}$ are the eigenvalues of the Laplace operator on the manifold and  $\{g_k\}$ are the dimensions of the corresponding eigenspaces.}.

To conclude this section, let us finally mention that, among the family of adiabatic states, those of sufficiently high order have the additional good physical property of allowing for a consistent definition of expectation values of the stress-energy tensor \cite{PF} (see also \cite{AAN}). 

\section{Rescaled field description and unitary dynamics}

Let us consider again the general case of an arbitrary scale factor $a(\eta)$, but now from the 
perspective of the rescaled field $\chi=a\phi$. Following again the strategy explained above, the natural way to proceed is to fix initial data
\be
\label{goodic}
\chi_k=\frac{1}{\sqrt{2\omega(\eta)}}, \qquad \chi'_k={-i}\sqrt{\frac{\omega(\eta)}{2}}, \qquad 
\omega^2(\eta) =k^2+m^2a^2(\eta)
\ee
at some  fixed instant $\eta$, thus defining a CS  ${\cal J}^{\chi}_{\eta}$. If we now compare the quantization obtained by applying this procedure at two different instants of time, say $\eta_1$ and $\eta_2$, it turns out that they are actually unitarily equivalent, i.e.,  ${\cal J}^{\chi}_{\eta_2}-{\cal J}^{\chi}_{\eta_1}$ is  Hilbert-Schmidt (see \cite{Fp}). Thus, the procedure that leads to ambiguities if the field $\phi$ is considered actually produces an unambiguous quantization when $\chi$ is used instead. Note that, as far as this aspect of the problem is concerned, the issue is not really about the field variable that is used: employing relation (\ref{qf}), one can certainly obtain now an unambiguous quantization procedure that works for the field $\phi$. What happens is that  the initial conditions corresponding to (\ref{goodic}) have to be used, and not (\ref{modein}) and (\ref{ome}). In fact, since the relation between $\phi$ and $\chi$ involves  explicitly the time,  the relation between Cauchy data at a given instant is
\be
\label{goodicphi}
\phi(\eta)=\frac{\chi(\eta)}{a(\eta)},
\ee
\be
\label{goodicphi2}
\phi'(\eta)=\frac{\chi'(\eta)}{a(\eta)}-\frac{a'(\eta)}{a^2(\eta)}\chi(\eta).
\ee
Therefore, the use of the rescaled field $\chi$ permits a natural implementation of the pursued quantization strategy, so that: 1) A physically motivated CS is introduced (like in the case of a constant scale factor in a given time interval), and 2) The procedure does not lead to ambiguities, what in the present case means that the unitary equivalence class of the representation that is determined in this way does not depend on the instant $\eta$ at which the initial conditions (\ref{goodic}) are imposed. We insist that we are not only obtaining a quantization for the field $\chi$: we are obviously defining a quantization of the field $\phi$, either directly  by $\hat \phi=a^{-1}\hat \chi$ or, equivalently, by means of the initial conditions which correspond to (\ref{goodic}), via (\ref{goodicphi}) and (\ref{goodicphi2}).

However, there is a major difference between the two field descriptions: whereas the time evolution of the field $\chi$ is unitary, that of $\phi$ is not. Let us see why the evolution of $\chi$ is indeed unitary. To help the discussion, let $J^{m(\eta_1)}$ denote the CS in $\Gamma_{\rm{Can}}$ defined by 
\be
J^{m(\eta_1)}=I_{\eta_1}{\cal J}^{\chi}_{\eta_1}I_{\eta_1}^{-1},
\ee
and similarly for 
\be
J^{m(\eta_2)}=I_{\eta_2}{\cal J}^{\chi}_{\eta_2}I_{\eta_2}^{-1}.
\ee
The above mentioned fact that ${\cal J}^{\chi}_{\eta_2}-{\cal J}^{\chi}_{\eta_1}$ is  Hilbert-Schmidt (proven e.g. in \cite{Fp}) means precisely that 
\be
\label{jev1}
E_{\eta_2,\eta_1} J^{m(\eta_1)} E^{-1}_{\eta_2,\eta_1}-J^{m(\eta_2)}
\ee
is  Hilbert-Schmidt. But it now happens that $J^{m(\eta_2)}-J^{m(\eta_1)}$ is Hilbert-Schmidt as well\footnote{This follows as well from Lemma 1 in \cite{AA}, with the obvious adaptations, because in the $\chi$-description the scale factor affects only the frequency, like in (\ref{ome}), but does not appear as a global factor in the initial data, as in (\ref{modein}).} (this is true only for compact spatial manifolds, see e.g. \cite{9}). So, we conclude then that 
\be
\label{jev2}
E_{\eta_2,\eta_1} J^{m(\eta_1)} E^{-1}_{\eta_2,\eta_1}-J^{m(\eta_1)}
\ee
is itself Hilbert-Schmidt, a result which tells us that the dynamics of the field $\chi$ is unitary, in the representation of the CCR's defined by $J^{m(\eta_1)}$.

From our perspective, unitary implementability of the dynamics is a most compeling mathematical physics requirement, and we advocate the use of a formulation which allows it, whenever possible. Moreover, in the present context of free fields in a (spatially) homogeneous and isotropic cosmological background, that requirement (together with invariance under spatial symmetries) is certainly a valid criterion to select the quantum representation. In this respect, let us recall that it was proven in \cite{9,10,15,16,18,21,22} that: i) Unitary implementability of the dynamics is impossible to achieve by means of any other rescaling different from $\phi \to \chi=a\phi$, including the formulation in terms of the original unscaled field $\phi$ (in this respect, see also \cite{Vi}); and ii) Once the formulation in terms of the field $\chi$ is chosen, there is a unique equivalence class of Fock representations such that a unitary implementation of the dynamics is possible (notice that, in proving these results, it is always assumed that the CS which defines the Fock quantization is invariant under spatial isometries; in  terms of the mode decomposition, this restriction is already encoded in the fact that the creation-annihilation operators do not mix different modes \cite{adia}). 

For completeness, let us also mention that the CS considered in the works \cite{9,10,15,16,18,21,22} (which allows for unitary dynamics) corresponds not exactly to the initial data (\ref{goodic}), but rather to
\be
\label{ouric}
\chi_k=\frac{1}{\sqrt{2 k}}, \qquad \chi'_k={-i}\sqrt{\frac{k}{2}}.
\ee
That the two sets of data (\ref{goodic}) and (\ref{ouric}) determine (for compact spatial manifolds) unitarily equivalent representations is easily proven directly, and was already mentioned in \cite{9}.

This all being said, let us analyze the relation between the quantum $\chi$-description, with unitary dynamics, and the   quantum $\phi$-description, determined by adiabatic states, which is the one favored e.g.~in \cite{AA}. First and foremost, the unique $\chi$-quantization with unitary dynamics, defined for instance by the data (\ref{ouric}) [or equivalently (\ref{goodic})] at some instant $\eta$, determines, via $\hat\phi=a^{-1}\hat\chi$, a quantum representation for $\phi$ which is unitarily equivalent to the one defined by the zeroth-order adiabatic state (this was proven explicitly in section VI of \cite{21} for the case of the 3-sphere, but again the proof is immediately applicable to the present 3-torus case, and hence we will not repeat it here). Since any two adiabatic states give rise to unitarily equivalent representations, independently of the order (as we showed in the previous section), we have that the quantization determined by the requirement of unitary dynamics is equivalent to the quantization defined by any adiabatic state.  Therefore, one does not lose the possibility of using adiabatic states by working in the Fock representation picked out by the data (\ref{ouric}) [or the corresponding data for $\phi$, via (\ref{goodicphi}) and (\ref{goodicphi2})]. The main difference is that the usual adiabatic state is no longer the {\it vacuum} of the Fock representation, but it is still a well-defined element of the Fock space determined by (\ref{ouric}) (in this sense, let us recall that another equivalent criterion for the unitary equivalence of two Fock representations is that the vacuum of one of the quantizations belongs to the Hilbert space of the other quantization, where the vacuum is defined as the state annihilated by all the annihilation operators). An explicit expression of that adiabatic state, for any order $r$, can be given as follows:
\be
\label{adi}
|0\rangle_{ad}^{(r)}= F^{(r)}\exp\left[-\frac{1}{2}\sum_{{\vec k}}\lambda^{(r)}_{k} \hat A^{\dagger}_{-\vec k}\hat A^{\dagger}_{\vec k}\right]|0\rangle,
\ee
where $|0\rangle$ is the vacuum associated with the initial conditions (\ref{ouric}), mapped to the $\phi$-field description via $\chi=a\phi$, and $\hat A^{\dagger}_{\vec k}$ are the corresponding creation operators. Here, $\lambda^{(r)}_k=\b^{(r)}_k/\a^{(r)}_k$, and $\a^{(r)}_k$ and $\b^{(r)}_k$ are the parameters of the Bogoliubov transformation that relates $|0\rangle_{ad}^{(r)}$ to $|0\rangle$ (defined of course at the same instant of time). Full expressions for these parameters can  be obtained following the analysis performed in section VI of \cite{21}. We get
\be
\alpha^{(r)}_{k}=-\frac{i a}{\sqrt{2k}}\left[\Big(\dot W^{(r)}_{k}\Big)^{*} a + \Big(W^{(r)}_{k}\Big)^{*}\left( \dot a +i k\right)\right],
\ee
\be
\beta^{(r)}_{k}=-\frac{i a}{\sqrt{2k}}\left[\Big(\dot W^{(r)}_{k}\Big)^{*} a + \Big(W^{(r)}_{k}\Big)^{*}\left(\dot a - i k \right)\right].
\ee
To facilitate comparison with the existing literature, we have now switched to cosmological time $\tau$, related with the conformal time $\eta$ by $d\tau=ad\eta$. Besides, the dot stands for the derivative with respect to the cosmological time, and the adiabatic mode solutions $W^{(r)}_{k}$ (constructed at fixed arbitrary time $\bar{\tau}$) have the expression
\begin{equation}
\label{ansatz-adiabatic}
W_{k}(\tau)=\frac{1}{\sqrt{2a^{3}\Omega_{k}}}
\exp\left(-i\int_{\bar{\tau}}^{\tau} \Omega_{k}(\tilde{\tau})d\tilde{\tau}\right).
\end{equation}
The requirement that these functions satisfy the equations of motion for the modes leads to the following equations for the time dependent frequencies $\Omega_{k}$ (see e.g. \cite{21}):
\begin{equation}
\label{lambda-sol}
\Omega^{2}_{k}=w_{k}^{2}-\frac{3}{4}\left(\frac{\dot a}{a}\right)^{2}
-\frac{3}{2}\frac{\ddot a}{a}+\frac{3}{4}\left(\frac{\dot\Omega_{k}}{\Omega_{k}}\right)^{2}
-\frac{1}{2}\frac{\ddot\Omega_{k}}{\Omega_{k}},
\end{equation}
where
\be
w^2_{k}=\frac{k^2}{a^{2}}+m^{2}.
\ee
The different adiabatic orders $r$ emerge when  approximate solutions to (\ref{lambda-sol}) are obtained  by means of an iterative process, where one obtains the $r$-th (positive) function $\Omega^{(r)}_{k}$ from the preceding one, $\Omega^{(r-1)}_{k}$:
\begin{equation}
\label{n-order}
\left(\Omega^{(r)}_{k}\right)^{2}=w_{k}^{2}-\frac{3}{4}\left(\frac{\dot a}{a}\right)^{2} -\frac{3}{2}\frac{\ddot a}{a}+\frac{3}{4}\left(\frac{\dot\Omega^{(r-1)}_{k}}{\Omega^{(r-1)}_{k}}\right)^{2} -\frac{1}{2}\frac{\ddot\Omega^{(r-1)}_{k}}{\Omega^{(r-1)}_{k}}, 
\end{equation}
with the initial condition
\be
\left(\Omega^{(0)}_{k}\right)^{2}=w_{k}^{2}.
\ee
Finally, in the expression of the adiabatic state, $F^{(r)}$ is a normalization factor, given by
\be
\left|F^{(r)}\right|=\prod_{\vec k}\left(1-\left|\lambda^{(r)}_k\right|^2\right)^{1/4}.
\ee

Hence, in particular, the usual regularization of, let's say, the stress-energy tensor can still be performed with respect to the above states (\ref{adi}) in the quantization that we have adopted.

\section{Discussion and conclusions}

In view of our previous discussion, we find no conflict or tension between the $\chi$-description based on the inital data (\ref{ouric}) and the $\phi$-description based e.g. on high-order adiabatic states (which is the description preferred, for instance, by the authors of \cite{AA}). As we have just shown, one can move freely from one description to the other; it is in fact the same quantization (up to unitary equivalence), with the difference that the dynamics of the field $\chi$ is unitary, while that of $\phi$ is not. Therefore, we see no need for a generalized notion of unitary  implementation of the dynamics.

From our perspective, a notion of generalized unitarity like that introduced in \cite{AA} is best seen as a consistency condition. Let us recall that, in that kind of approach, one is allowed to work {in practice} with a family of CS's $J_{\eta}$ in $\Gamma_{\rm{Can}}$, and that this family is said to lead to a generalized unitary implementation of the dynamics if 
\be
\label{AAuni}
J_{\eta_2} - E_{\eta_2,\eta_1} J_{\eta_1} E^{-1}_{\eta_2,\eta_1}
\ee
is  Hilbert-Schmidt for all possible instants of time $\eta_1$ and $\eta_2$ (it is worth noticing that this family is now not even necessarily tied to a single CS in $\Gamma_{\rm{Cov}}$). It is absolutely true that if one tries to quantize a linear field following this route, one should require the above condition as a minimal consistency condition: it is clear that otherwise one would not get a well-defined quantization. In fact, from the family of CS's in $\Gamma_{\rm{Can}}$ one can construct, via  (\ref{7}), a family of CS's in $\Gamma_{\rm{Cov}}$. The discussed condition (\ref{AAuni}) is precisely what guarantees that all those CS's in $\Gamma_{\rm{Cov}}$ lead to unitarily equivalent representations. However, as far as we can understand, the condition guarantees nothing more than that, and in particular it is not clear at all how a specific quantization appropriate for a given field theory is selected in that way. In this respect, note that any quantization defined by a single CS $\cal J$ in $\Gamma_{\rm{Cov}}$ satisfies the condition trivially, since the family obtained via (\ref{7}) gives exactly zero for all operators of the form (\ref{AAuni}) obtained for different pairs of instants $\eta_1$ and $\eta_2$.

Certainly, the S-matrix fits in the proposed generalized unitarity scheme (as pointed out in \cite{AA}). In fact, as we have seen in situations with a constant scale factor in the far past and future, the difference of CS's  ${\cal J}_{in}-{\cal J}_{out}$ is  Hilbert-Schmidt, or equivalently $J_{out} - E_{\eta_2,\eta_1} J_{in} E^{-1}_{\eta_2,\eta_1}$ is Hilbert-Schmidt, but again this is a consistency condition: the CS's ${\cal J}_{in}$ and ${\cal J}_{out}$ themselves are selected by additional physical requirements, in this case by matching the CS with that of a free field in regions where the spacetime is isomorphic to Minkowski spacetime. So, additional mathematical physics requirements, such as those that we have just commented, or the ultraviolet regularity advocated in \cite{AA}, or,  indeed, a unitary implementation of the dynamics, are required in order to pick out an appropriate quantization. Besides, it is worth noticing that the unitary implementation of the dynamics is actually a condition of the same nature of ultraviolet regularity, in the sense that, like the requirement of adiabatic regularity, it does control the ultraviolet behavior. In fact, the requirement of unitary dynamics leads to a class of CS's in which  the ultraviolet properties are essentialy fixed, in the sense that the difference between two such CS's is always a Hilbert-Schmidt operator. 

To conclude our discussion, let us just mention that, while it is true that once one has a quantization with unitary dynamics for the rescaled field $\chi$, one can always obtain a quantization for the field $\phi$ which satisfies the generalized unitarity condition (as shown in \cite{AA}), there is no guarantee that the converse is true. So, the condition of unitary implementation of the dynamics and the generalized unitarity condition are not equivalent. 

\section*{Acknowledgements}

We acknowledge financial aid from the research grants MICINN/MINECO FIS2011-30145-C03-02 and FIS2014-54800-C2-2-P from Spain,
 DGAPA-UNAM IN113115 and CONACyT 237351 from Mexico. The authors are thankful to I. Agullo for correspondence.


\begin{thebibliography}{99}


\bibitem{9} A. Corichi, J. Cortez, G.A. Mena Marug\'an, and J.M. Velhinho,
Class. Quantum Grav. \textbf{23}, 6301 (2006).

\bibitem{10} J. Cortez, G.A. Mena Marug\'an, and J.M. Velhinho, Phys. Rev. D \textbf{75}, 084027 (2007).

\bibitem{15} J. Cortez, G.A. Mena Marug\'an, R. Ser\^odio, and J.M. Velhinho, Phys. Rev. D \textbf{79}, 084040 (2009).

\bibitem{16} J. Cortez, G.A. Mena Marug\'an, and J.M. Velhinho, Phys. Rev. D \textbf{81}, 044037 (2010).

\bibitem{CQG28} J.~Cortez, G.~A.~Mena Marug\'{a}n, J.~Olmedo, and J.~M.~Velhinho,  Class.\ Quantum Grav.\  {\bf 28}, 172001 (2011).


\bibitem{21} J. Cortez, G.A. Mena Marug\'an, J. Olmedo, and J.M. Velhinho, Phys. Rev. D \textbf{86}, 104003 (2012).

\bibitem{18} J. Cortez, G.A. Mena Marug\'an, J. Olmedo, and J.M. Velhinho, Phys. Rev. D \textbf{83}, 025002 (2011).

\bibitem{22} L. Castell\'o Gomar, J. Cortez, D. Mart\'\i n-de Blas, G.A. Mena Marug\'an, and J.M. Velhinho, JCAP \textbf{11}, 001 (2012).

\bibitem{Fb} S.A. Fulling, {\it Aspects of Quantum Field Theory in Curved Spacetime} (Cambridge University Press, Cambridge, 1989).
 
\bibitem{BD} N.D. Birrell and P.C.W. Davies, {\it Quantum Fields in Curved Space} (Cambridge University Press, Cambridge, 1982).

\bibitem{AA} A. Ashtekar and I. Agullo, 
Phys. Rev. D \textbf{91}, 124010 (2015).

\bibitem{wald} R.M. Wald, {\it Quantum Field Theory in Curved Spacetime and Black Hole Thermodynamics} (Chicago University Press, Chicago, 1994).

\bibitem{adia} C. L\"{u}ders and J.E. Roberts, Commun. Math. Phys. \textbf{134}, 29 (1990).

\bibitem{Fp} S.A. Fulling, Gen. Rel. Grav. \textbf{10}, 807 (1979).

\bibitem{kay} B.S. Kay, Commun. Math. Phys. \textbf{62}, 55 (1978).

\bibitem{BSZ} J.C. Baez, I.E. Segal, and Z. Zhou, {\it Introduction to Algebraic and Constructive Quantum Field Theory} (Princeton University Press, Princeton, 1992).

\bibitem{AM1} A. Ashtekar and A. Magnon, Proc. R. Soc. (London) A \textbf{346}, 375 (1975).
 
 \bibitem{AM2}A. Ashtekar and A. Magnon-Ashtekar, Pramana \textbf{15}, 107 (1980).

\bibitem{jackiw} R. Floreanini, C.T. Hill, and R. Jackiw, Ann. Phys. \textbf{175}, 345 (1987).

\bibitem{W} R.M. Wald, 
Ann. Phys. \textbf{118}, 490 (1979).

\bibitem{D} J. Dimock, 
J. Math. Phys. \textbf{20}, 2549 (1979).

\bibitem{Par1}  L. Parker, Phys. Rev. Lett. \textbf{21}, 562 (1968).


\bibitem{Par2}  L. Parker, Phys. Rev. \textbf{183}, 1057 (1969).

\bibitem{PF} L. Parker and S. A. Fulling, 
Phys. Rev. D \textbf{9}, 341 (1974).

\bibitem{AAN} I. Agullo, A. Ashtekar, and W. Nelson, 
Class. Quantum Grav. \textbf{30}, 085014 (2013).

\bibitem{Vi} S. Vitenti, {\it Unitary evolution, canonical variables and vacuum choice for general quadratic Hamiltonians in spatially homogeneous and isotropic space-times}, arXiv:1505.01541.



\end{thebibliography}
\end{document}